\documentclass[twoside,11pt]{article}

\usepackage{jmlr/jmlr2e}

\usepackage[english]{babel}
\usepackage{natbib}
\usepackage[letterpaper,top=2cm,bottom=2cm,left=3cm,right=3cm,marginparwidth=1.75cm]{geometry}

\usepackage{csquotes}

\usepackage{amsmath}
\usepackage{amsfonts}
\usepackage{graphicx}

\usepackage{algorithm} 
\usepackage{algpseudocode} 
\usepackage{algcompatible}
\algnewcommand\INPUT{\item[\textbf{Input:}]}%
\algnewcommand\OUTPUT{\item[\textbf{Output:}]}%

\algdef{SE}[SUBALG]{Indent}{EndIndent}{}{\algorithmicend\ }%
\algtext*{Indent}
\algtext*{EndIndent}

\def\wt{\widetilde}

\def\wh{\widehat}

\newcommand{\beq}{\begin{equation}}
	\newcommand{\eeq}{\end{equation}}
\newcommand{\bea}{\begin{eqnarray}}
	\newcommand{\eea}{\end{eqnarray}}
\newcommand{\beas}{\begin{eqnarray*}}
	\newcommand{\eeas}{\end{eqnarray*}}

\newcommand{\bct}{\begin{center}}
	\newcommand{\ect}{\end{center}}

\newcommand{\bA}{ {\bf A} }

\newcommand{\bC}{ {\bf C} }

\newcommand{\bR}{ {\bf R} }

\newcommand{\bU}{ {\bf U} }

\newcommand{\bX}{ {\bf X} }
\newcommand{\by}{ {\bf y} }

\newcommand{\bbeta}{ {\boldsymbol \beta} }

\newcommand{\bGamma}{ {\boldsymbol \Gamma} }

\newcommand{\bmu}{ {\boldsymbol \mu} }

\newcommand{\bSigma}{ {\boldsymbol \Sigma} }

\makeatletter
\newcommand*\bigcdot{\mathpalette\bigcdot@{.5}}
\newcommand*\bigcdot@[2]{\mathbin{\vcenter{\hbox{\scalebox{#2}{$\m@th#1\bullet$}}}}}
\makeatother

\def\bbeta{\boldsymbol{\beta}}

\def\bGamma{\boldsymbol{\Gamma}}

\def\pr{\mathrm{P}}

\def\bR{\mathbf{R}}

\def\bSigma{\boldsymbol{\Sigma}}

\def\bU{\mathbf{U}}

\def\bX{\mathbf{X}}

\usepackage{lastpage}
\jmlrheading{VOLUME}{YEAR}{1-\pageref{LastPage}}{DATE ; Revised DATE}{PUBLISHED}{ID}{Evan Mason and Zhe Fei}

\ShortHeadings{Heterogeneous Nonparametric Knockoffs}{Mason and Fei}
\firstpageno{1}

\begin{document}
\title{Novel Knockoff Generation and Importance Measures with Heterogeneous Data via Conditional Residuals and Local Gradients}

\author{\name Evan Mason \email emaso007@ucr.edu \\
    \addr Department of Statistics\\
    University of California, Riverside\\
    Riverside, CA 92521
    \AND
    \name Zhe Fei \email zhef@ucr.edu \\
    \addr Department of Statistics\\
    University of California, Riverside\\
    Riverside, CA 92521}
\editor{My Editor}

\maketitle

\begin{abstract}
Knockoff variable selection is a powerful framework that creates synthetic “knockoff” variables to mirror the correlation structure of the observed features, enabling principled control of the false discovery rate in variable selection. However, existing methods often assume homogeneous data types or known distributions, limiting their applicability in real-world settings with heterogeneous, distribution-free data. Moreover, common variable importance measures rely on linear outcome models, hindering their effectiveness for complex relationships.
We propose a flexible knockoff generation framework based on conditional residuals that accommodates mixed data types without assuming known distributions. To assess variable importance, we introduce the Mean Absolute Local Derivative (MALD), an interpretable metric compatible with nonlinear outcome functions, including random forests and neural networks. Simulations show that our approach achieves better false discovery rate control and higher power than existing methods. We demonstrate its practical utility on a DNA methylation dataset from mouse tissues, identifying CpG sites linked to aging. Software is available in R (rangerKnockoff
) and Python (MALDimportance).
\end{abstract}





\section{Introduction}

In high-dimensional settings such as genomics, economics, and finance, identifying the most relevant predictors is essential for both accurate modeling and scientific insight. A key challenge in these domains is to control the false discovery rate (FDR) while selecting variables from a large pool of potential predictors, often with sample sizes much smaller than the number of features.

The knockoffs framework provides a powerful solution by creating synthetic “knockoff” variables that mimic the correlation structure of the original predictors. In essence, knockoffs serve as negative controls that allow one to compare the importance of each original variable against its knockoff copy. Early work \cite{barber2015controlling,candes2018panning} introduced knockoffs in the Gaussian setting via moment matching—often referred to as second-order knockoffs—and later the Model-X knockoffs framework extended the idea to general high-dimensional nonlinear models \cite{candes2018panning}. More recently, methods such as Deep Knockoffs \cite{romano2020deep} have leveraged deep generative models to approximate the knockoff distribution under less restrictive assumptions.

Despite these advancements, current methods face important limitations. Many existing procedures assume homogeneous data (e.g., all predictors are continuous and follow a known distribution) and rely on second-order moment matching. Such assumptions can be restrictive in real-world applications where data are heterogeneous—often including mixtures of categorical and continuous variables—and the relationship between predictors and the outcome may be nonlinear.

In this paper, we propose new methods that overcome these limitations by:
\begin{itemize}
    \item \textbf{Handling heterogeneous data:} We develop a knockoff generation procedure based on conditional residuals that adapts to unknown and mixed data types. By avoiding strong distributional assumptions, our method is applicable to settings where predictors exhibit diverse scales and types.
    \item \textbf{Accommodating nonlinear outcomes:} We introduce a novel, interpretable importance measure called Mean Absolute Local Derivative (MALD) that captures variable effects in complex, nonlinear outcome models. This measure allows for flexible inference without relying on a prespecified outcome model.
\end{itemize}

Our approach builds on the extensive literature on knockoffs while addressing their limitations in heterogeneous and nonlinear settings. The remainder of the paper is organized as follows. In Section 2, we review existing literature and summarize the classical knockoffs methodology. Section 3 presents our new methods for knockoff generation and variable importance in the context of heterogeneous data and nonlinear outcomes. Theoretical properties—including asymptotic FDR control and power analysis—are established in Section 4. Section 5 illustrates the finite-sample performance of our approach through simulation studies, and Section 6 applies our methods to real data. Finally, Section 7 concludes with a discussion of potential extensions and future research directions.

\section{Existing Literature}
\label{sec:literature}

The literature on knockoff methods spans a broad spectrum—from original approaches assuming Gaussian predictors to modern nonparametric and deep learning techniques, as well as methods tailored for categorical and heterogeneous data. In what follows, we review these developments by grouping them into three main themes.

\subsection{Knockoffs for Gaussian Data and Summary Statistics}

For Gaussian predictors, knockoff methods have been rigorously developed. In \cite{barber2015controlling,candes2018panning}, the concept of second-order knockoffs is introduced by requiring that the joint vectors $[X,\wt{X}]$ and their swapped versions $[X,\wt{X}]_{\text{swap}(S)}$ share the same first two moments. Specifically, the covariance matrix of the joint vector is defined as
$$
    G = \begin{bmatrix}
        \Sigma & \Sigma - \text{diag}\{s\} \\
        \Sigma - \text{diag}\{s\} & \Sigma
    \end{bmatrix},
$$
where $D = \text{diag}(s_1,\ldots,s_p)$ is computed by solving the convex optimization problem: 
$$
    \text{minimize}\ \sum_{j}|1 - s_j|,\quad \text{subject to}\quad 
    \begin{cases}
        2\Sigma - D \succeq 0, \\
        s_j \geq 0, \quad 1\leq j\leq p.
    \end{cases}
$$

\subsection{Neural Network-Based and Nonparametric Approaches}

To overcome the limitations imposed by assuming known distributions, several methods have employed neural networks for nonparametric knockoff generation. For instance, \cite{jordon2018knockoffgan} leverages a Generative Adversarial Network (GAN) to create knockoffs that replicate the original data distribution while incorporating a decorrelation penalty through a Mutual Information Neural Network \citep{belghazi2018mutual}. In a similar vein, \cite{romano2020deep} proposes a method that optimizes a weighted objective function combining maximum mean discrepancy, second moment matching, and a decorrelation term to reduce the correlation between each variable $X_j$ and its knockoff $\wt{X}_j$. Alternatively, \cite{liu2018auto} uses an autoencoder to generate knockoffs that preserve the encoded representation, which is closely related to the IPAD method introduced in \cite{fan2020ipad}.

Despite their flexibility, neural network-based approaches have two main drawbacks: they are often computationally intensive with extensive hyperparameter tuning, and their efficacy with heterogeneous data remains largely untested. 

\subsection{Methods for Categorical and Heterogeneous Data}

While knockoffs assuming known distributions yield strong analytic results, they are sometimes too restrictive for practical, heterogeneous data. Several works have addressed this gap by focusing on categorical and mixed data types. For example:
\begin{itemize}
    \item \cite{martens2021bayesian} introduces a Bayesian framework that employs a Gaussian Dirichlet Process Mixture (GDPM) to model continuous variables, along with latent log-likelihoods for categorical variables.
    \item \cite{kormaksson2020sequential} uses the Sequential Conditional Independent Pairs (SCIP) method, modeling continuous variables via linear functions and incorporating log-likelihoods for categorical variables.
    \item \cite{blesch2024conditional} proposes a streamlined SCIP model that generates both continuous and categorical variables by linking categorical outcomes to quantiles of a latent normal distribution, demonstrating improved performance over deep knockoff approaches in certain cases.
    \item \cite{sesia2019gene} employs hidden Markov models for generating multilevel categorical variables, while \cite{bates2021metropolized} tackles categorical data within a known model framework using MCMC techniques, though at the cost of increased implementation complexity.
\end{itemize}
These developments highlight the ongoing evolution of knockoff methods toward greater flexibility and applicability in real-world settings, especially for heterogeneous data types.

In summary, while classical knockoff methods for Gaussian data have provided a strong theoretical foundation, recent advances—including summary statistics extensions, nonparametric approaches using neural networks, and specialized methods for categorical data—underscore the need for flexible, computationally efficient frameworks capable of handling the complexity of modern datasets.

\section{Proposed Conditional Residual Knockoffs}\label{section:proposed-0}

The key novelty of our proposed method is to generate knockoffs using the conditional residuals of the variables by removing the conditional expectations. Advantages of this approach include: i) it is more robust to variable dependency; ii) it can capture nonlinear relationships and high-order dependencies; iii) it better approximates true heterogeneous covariate distribution.

In this work, the observed data matrix is denoted by $\bX \in \mathbb{R}^{n \times p}$, where each row $\bX_{i\cdot} \in \mathbb{R}^{p}$ represents the $i$th observation, and each column corresponds to one variable. The knockoff data matrix is denoted by $\wt{\bX} \in \mathbb{R}^{n \times p}$. For any given observation $\bX_{i\cdot}$, we use $\bX_{i,-j}$ to denote the subvector of $\bX_{i\cdot}$ excluding the $j$th variable. We further define the predictor $\wh{\bX}_{i\cdot}$ of the conditional expectation as
$\wh{\bX}_{i\cdot} = \left(\wh{\bX}_{i1},\wh{\bX}_{i2},\ldots,
    \wh{\bX}_{ip} \right)$,  $\wh{\bX}_{ij} = \wh{g}_j(\bX_{i,-j})$,
where $\wh{g}_j: \mathbb{R}^{p-1} \rightarrow \mathbb{R}$ is an estimator of the conditional expectation $\mathbb{E}[\bX_{ij} \mid \bX_{i,-j}]$. Examples of $\wh{g}_j$ include the node-wise Lasso \citep{meinshausen2006high}, and random forests for non-linear modeling.
Then the {\bf conditional residuals} $\Gamma(\bX_{i\cdot})$ are defined as $\Gamma(\bX_{i\cdot}) = \bX_{i\cdot} - \wh{\bX}_{i\cdot}.$
The key idea is to apply existing knockoff methods to $\Gamma(\bX_{i\cdot})$'s and generate valid knockoffs denoted as $\wt{\Gamma}(\bX_{i\cdot})$'s.

To handle mixed data with categorical variables, we extend the conditional residual definition as follows. 
Let $J_C \subset \{1,\ldots,p\}$ denote the set of categorical variable indices. We first define an intermediate matrix $\bU \in \mathbb{R}^{n \times p}$:
\[
    \bU_{ij} = \begin{cases}
        0, & \text{if } j \in J_C,\\[1mm]
        \wh{\bX}_{ij}, & \text{if } j \notin J_C.
    \end{cases}
\]
For each categorical feature $j \in J_C$, $X_{ij} \in \{1,\ldots, K_j\}$, where $K_j$ is the number of classes. Therefore, $\wh{g}_j: \mathbb{R}^{p-1} \rightarrow \{1,\ldots, K_j\}$ becomes an estimator of the conditional probabilities $\pr \left( X_{ij} = k \mid \bX_{i,-j} \right), k \in \{1,\ldots, K_j\}$.
And we generate the knockoff residuals $\wt{\Gamma}_{ij}$ by sampling from the estimated conditional distribution:
\begin{equation}\label{cond_p}
    \pr\left( \wt{\Gamma}_{ij} = k \mid \bX_{i,-j} \right) = \pr\left( X_{ij} = k \mid \bX_{i,-j} \right), \quad \text{for } k \in \{1,\ldots, K_j\}.
\end{equation}
Lastly, we construct the knockoff feature matrix $\wt{\bX} \in \mathbb{R}^{n \times p}$ by combining the conditional expectation estimates and the knockoff residuals:
\begin{equation}\label{CR_knockoffs}
     \wt{\bX}_{i\cdot} = \bU_{i\cdot} + \wt{\Gamma}(\bX_{i\cdot}).
\end{equation}
The complete CR knockoff procedure is described in Algorithm \ref{alg:conditional-residual-estimator-heterogeneous-0}.
Next, we show that the exchangeability still holds for conditional residual knockoffs, i.e. the procedure generates valid knockoffs.

\begin{theorem}[Exchangeability of Conditional Residual Knockoffs]\label{thm1}
Let $X = (X_1,\ldots, X_p) \in \mathbb{R}^p$ be a random vector, and the observed data are $\bX \in \mathbb{R}^{n \times p}$. Denote term-wise conditional expectation predictors $ \wh{\bX}_{\cdot j} = \wh{g}_j\left(\bX_{\cdot, -j} \right)$, $j=1,2,\ldots,p$ fit on the observed data $\bX$, and assume that $\wh{g}_j$ is an unbiased estimator of $\mathbb{E}[X_j \mid X_{-j}]$. Define the residual vectors $\bGamma_{\cdot j} = \bX_{\cdot j} - \wh{\bX}_{\cdot j}$. Suppose we can generate knockoff residuals $\wt{\bGamma}_{\cdot j}$ such that
\[
    (\bGamma, \wt{\bGamma}) \stackrel{d}{=} (\bGamma, \wt{\bGamma})_{\text{swap}(S)}, \quad \forall S \subseteq \{1,\ldots,p\}.
\]
Then, defining knockoff variables by $\wt{\bX} = \wh{\bX} + \wt{\bGamma}$,
it follows that
\[
    (\bX, \wt{\bX}) \stackrel{d}{=} (\bX, \wt{\bX})_{\text{swap}(S)}, \quad \forall S \subseteq \{1,\ldots,p\}.
\]
\end{theorem}

The proof is deferred to the Appendix. Concretely, we propose the forest conditional residual knockoffs (Forest CRK) using random forests to fit each $g_j$, so that for numeric variables $j\not\in J_C$, $ U_{ij} = g_j(\bX_{i,-j}).$ For categorical variables $j \in J_C$, the random forest would fit the conditional probabilities as in (\ref{cond_p}).

\begin{algorithm}
	\caption{Conditional Residual Forest with Heterogeneous Data
        \label{alg:conditional-residual-estimator-heterogeneous-0}
    } 
	\begin{algorithmic}[1]
        \REQUIRE Predictive model $\Psi$ (e.g., random forest), knockoff generator $C$ 
        \INPUT Observed data $\bX\in \mathbb{R}^{n\times p}$ 
        \OUTPUT Knockoff matrix $\wt{\bX}$
                
        \State Initialize predicted matrix $\wh{\bX} \leftarrow 0$, residual matrix $\wh{\bGamma} \leftarrow 0$
        \FOR{$j = 1$ to $p$}
            \State Fit $\wh{g}_j = \Psi(\bX_{\cdot j}, \bX_{\cdot, -j})$
            \IF{$\bX_{\cdot j}$ is numeric}
                \State $\wh{\bX}_{ij} \gets \wh{g}_j(\bX_{i, -j})$
                \State $\wh{\bGamma}_{ij} \gets \bX_{ij} - \wh{\bX}_{ij}$
            \ELSE \Comment{Categorical}
                \State Store class probabilities $\wh{g}_j(k \mid \bX_{i, -j})$ for all $k \in \{1,\ldots,K_j\}$
                \State Set $\wh{\bX}_{ij} \gets 0$ and define residuals implicitly
            \ENDIF
        \ENDFOR
            
        \State Generate knockoffs of residuals: $\wt{\bGamma}_{i \cdot} \gets C(\wh{\bGamma}_{i \cdot})$
        
        \FOR{$j = 1$ to $p$}
        \IF{$\bX_{\cdot j}$ is numeric}
            \State $\wt{\bX}_{ij} \gets \wh{\bX}_{ij} + \wt{\bGamma}_{ij}$
        \ELSE \Comment{Categorical}
            \State Sample $\wt{\bX}_{ij}$ from categorical distribution $\propto \wh{g}_j(k \mid \bX_{i, -j})$ for $k = 1,\ldots,K_j$
        \ENDIF
        \ENDFOR
        
        \State Return $\wt{\bX}$
	\end{algorithmic} 
\end{algorithm}

\subsection{Mixture Gaussian Distribution as an Example}

Suppose $\bX_{i\cdot} \sim \mathcal{N}(\bmu, \bSigma)$. By the formula for the conditional distribution of multivariate Gaussian, the conditional expectation of $X_{ij}$ given the other variables $\bX_{i,-j}$ is \(
    \mathbb{E}[X_{ij} \mid \bX_{i,-j}] = \mu_j + \bSigma_{j,-j} \bSigma_{-j,-j}^{-1}(\bX_{i,-j} - \bmu_{-j}).
\)
This expression is an affine transformation of $\bX_{i,-j}$ and can be rearranged as:
\[
\begin{aligned}
    \mathbb{E}[X_{ij} \mid \bX_{i,-j}]
    &= \mu_j - \bSigma_{j,-j} \bSigma_{-j,-j}^{-1} \bmu_{-j} + \bSigma_{j,-j} \bSigma_{-j,-j}^{-1} \bX_{i,-j} \\
    &= \left(\mu_j - \bSigma_{j,-j} \bSigma_{-j,-j}^{-1} \bmu_{-j}\right) + \bX_{i,-j}^\top \left(\bSigma_{j,-j} \bSigma_{-j,-j}^{-1} \right).
\end{aligned}
\]
The residual for variable $j$ is then:
\[
\begin{aligned}
    X_{ij} - \mathbb{E}[X_{ij} \mid \bX_{i,-j}]
    &= X_{ij} - \left(\mu_j - \bSigma_{j,-j} \bSigma_{-j,-j}^{-1} \bmu_{-j} + \bX_{i,-j}^\top \bSigma_{j,-j} \bSigma_{-j,-j}^{-1} \right) \\
    &= (X_{ij} - \mu_j) + \bSigma_{j,-j} \bSigma_{-j,-j}^{-1} (\bmu_{-j} - \bX_{i,-j}) \\
    &= \left[1, -\bSigma_{j,-j} \bSigma_{-j,-j}^{-1} \right] \bX_{i\cdot} + \left( \bSigma_{j,-j} \bSigma_{-j,-j}^{-1} \bmu_{-j} - \mu_j \right).
\end{aligned}
\]
Repeating this for all $j = 1, \ldots, p$, the vector of residuals becomes \(
    \bX_{i\cdot} - \mathbb{E}[\bX_{i\cdot} \mid \bX_{i,-}] = \bC + \bA \bX_{i\cdot}\),
where \(\bC_j = \bSigma_{j,-j} \bSigma_{-j,-j}^{-1} \bmu_{-j} - \mu_j\),
   \(\bA_{j,\cdot} = \left[ 1, -\bSigma_{j,-j} \bSigma_{-j,-j}^{-1} \right],\)
and thus \(
    \bX_{i\cdot} - \mathbb{E}[\bX_{i\cdot} \mid \bX_{i,-}] \sim \mathcal{N}(0, \bA \bSigma \bA^\top).
\)
This shows that under the Gaussian assumption, the residuals from the conditional expectations are themselves normally distributed, and knockoffs for the residuals can be generated using standard methods such as the second-order construction via convex optimization \citep{candes2018panning}, or other approaches such as those proposed in \citep{ke2024power}.

\textbf{Residuals from Mixture Distributions.} Consider a mixture distribution where each sample $\bX_{i\cdot}$ is drawn from one of several modes indexed by $m$, i.e.:
\[
    \bX_{i\cdot} \sim \sum_{m} P(M_i = m) \cdot \mathbb{P}(\bX_{i\cdot} \mid M_i = m).
\]
Assume that the conditional residual covariance $\text{Cov}(\bGamma_{i\cdot} \mid M_i = m)$ is the same across $m$. Then, by the law of total variance:
\[
\begin{aligned}
    \text{Cov}(\bGamma_{i\cdot})
    &= \text{Cov}\left( \mathbb{E}[\bGamma_{i\cdot} \mid M_i] \right)
    + \mathbb{E} \left[ \text{Cov}(\bGamma_{i\cdot} \mid M_i) \right] \\
    &= \text{Cov}(0) + \sum_m P(M_i = m) \cdot \text{Cov}(\bGamma_{i\cdot} \mid M_i = m) \\
    &= \text{Cov}(\bGamma_{i\cdot} \mid M_i = m).
\end{aligned}
\]
Thus, even in mixture settings, the residuals can remain homoskedastic and mean-zero—maintaining properties desirable for knockoff construction.

\subsection{Sequential Conditional Independent Pairs with Categorical Variables}

The Sequential Conditional Independent Pairs (SCIP) framework provides a flexible approach for generating knockoffs under arbitrary data distributions. It has been applied in a variety of contexts, including heterogeneous datasets, using linear models with shrinkage or regularization \citep{kormaksson2020sequential,blesch2024conditional}. These approaches are most effective when the conditional models closely approximate the true data-generating process.

For Gaussian data, SCIP is equivalent to the second-order knockoff construction, which implicitly assumes that each variable can be modeled as a linear function of the others. However, in situations where the distribution of $\bX$ is unknown or not well approximated by linear models, nonparametric methods such as random forests can be used to consistently estimate the conditional expectations required for SCIP knockoff generation. 

\paragraph{Linear SCIP for Continuous Features.} In the standard linear SCIP framework, each continuous variable $\wt{\bX}_{\cdot j}$ is generated using a linear model trained on the observed variables and previously sampled knockoffs:
\[
\begin{aligned}
    \wh{g}_j([\bX_{i,-j}, \wt{\bX}_{i,1:j-1}]) &\approx \mathbb{E}[X_{ij} \mid \bX_{i,-j}, \wt{\bX}_{i,1:j-1}] \\
    &= [\bX_{i,-j}, \wt{\bX}_{i,1:j-1}] \wh{\bbeta}^{(j)}, \\
    \wt{\bX}_{ij} &= [\bX_{i,-j}, \wt{\bX}_{i,1:j-1}] \wh{\bbeta}^{(j)} + \wt{\epsilon}_{ij}, \\
    \wt{\epsilon}_{ij} &\overset{\text{iid}}{\sim} \mathcal{N}(0, \wh{s}_j^2), \\
    \wh{s}_j^2 &= \frac{1}{n-1} \sum_{i=1}^n \left( X_{ij} - [\bX_{i,-j}, \wt{\bX}_{i,1:j-1}] \wh{\bbeta}^{(j)} \right)^2.
\end{aligned}
\label{equation:scip-0}
\]

\paragraph{Extension to Categorical Features.} For categorical variables $\bX_{\cdot j}$, a common approach is to fit a multinomial logistic regression model to $[\bX_{i,-j}, \wt{\bX}_{i,1:j-1}]$, and then sample $\wt{X}_{ij}$ from the resulting probability distribution.

Alternatively, to accommodate heterogeneous or nonlinear data structures, one can use random forests to fit conditional models, i.e. $\wh{g}_j([\bX_{i,-j}, \wt{\bX}_{i,1:j-1}])$ is the random forest model between $\bX_j$ and $[\bX_{i,-j}, \wt{\bX}_{i,1:j-1}]$. Natually they can estimate conditional class probabilities for categorical features, allowing sampling of $\wt{X}_{ij}$ directly from the fitted random forest.

\paragraph{Trade-offs.} If the true conditional expectation is close to linear, or if the variable importance metric depends only on first and second moments, then a linear SCIP model may offer better power and efficiency. In such cases, the flexibility of random forests may be unnecessary or even detrimental due to higher variance in estimation.

However, for complex, nonlinear, or heterogeneous data—as is often encountered in modern applications—random forest-based SCIP knockoffs can provide a more robust alternative. This approach is especially useful when paired with nonparametric outcome models or when the assumptions of traditional Second-order knockoffs are violated. 

\section{Mean Absolute Local Derivative (MALD) as Importance Measures}

Most knockoff procedures to date assume a linear outcome model between $\by$ and $\bX$:
\[
    \mathbb{E}[y_i \mid \bX_{i\cdot}] = \bX_{i\cdot} \bbeta.
\]
The most common importance measure in this setting are based on LASSO coefficients, namely Lasso Coefficient Difference (LCD) and Lasso–Max \citep{weinstein2017power,weinstein2020power}. LCD is defined as follows:
\begin{enumerate}
    \item Fit a LASSO on the augmented dataset $[X, \tilde{X}]$, tuning the regularization parameter $\lambda$ (often via cross-validation). Denote the LASSO coefficients as $\wh{\beta} = \left( \wh{\beta}^{\text{orig}}, \wh{\beta}^{\text{knock}}\right)$.
    
    \item Obtain coefficient estimates: \( T_j = \wh{\beta}_j^{\text{orig}}(\lambda), \quad
        \wt{T}_j = \wh{\beta}_j^{\text{knock}}(\lambda).\)

    \item Compute the knockoff statistic:\(  W_j = |T_j| - |\wt{T}_j|.\)
\end{enumerate}
A larger positive $W_j$ suggests that the original feature $j$ is more predictive than its knockoff, making it a strong candidate for selection. On the other hand, the Lasso-Max method defines $T_j = \sup\{\lambda: \wh{\beta}_j(\lambda) \neq 0\}$, $\wt{T}_j = \sup\{\lambda: \wt{\beta}_j(\lambda) \neq 0\}$ and $W_j = T_j - \wt{T}_j$. Here $T_j$ and $\wt{T}_j$ are the respective largest penalty at which the coefficient for (knockoff) feature $j$ remains nonzero.

LASSO is particularly appealing for high-dimensional data and has been extensively studied, including analytic characterizations of power in terms of sample size, signal strength, and predictor correlation \citep{ke2024power}.
LASSO-based measures work well under linear or approximately linear outcomes, including when categorical variables are one-hot encoded. In that case, a categorical variable $j$ receives its own coefficient $\beta_{j_k}$ among the $k = 1,\ldots,K_j$ levels. 

However, when the outcome model is nonlinear or unknown, few importance measures perform reliably across heterogeneous data types. In the context of random forests, the change in Gini impurity—often used as an importance measure—has well-documented limitations, particularly for categorical variables. For instance, \citet{nicodemus2011stability} notes:
\begin{displayquote}
    MDG [mean decrease in Gini] has been shown to be sensitive to predictors with different scales of measurement (e.g., binary versus continuous) and shows artificial inflation for predictors with larger numbers of categories.
\end{displayquote}
Similar criticisms appear in \citet{strobl2007bias, gregorutti2017correlation}.

To address these issues, we propose the \textbf{Mean Absolute Local Derivative (MALD)}, a general-purpose importance measure that accommodates heterogeneous predictors and nonlinear outcome models. MALD relies on estimating the gradient of the fitted function $\mathbb{E}[\by_i \mid \bX_{i\cdot}]$ at each sample point $\bX_{i\cdot}$. Since local gradients can change sign (e.g., due to quadratic effects), we take the absolute value to avoid cancellation. For a chosen power $r > 0$, the MALD score for variable $j$ is defined as:
\[
    T_j = \frac{1}{n} \sum_{i=1}^n \left| \ell_{ij} \right|^r = \frac{1}{n} \sum_{i=1}^n \left| \frac{\partial}{\partial X_{ij}} \mathbb{E}[\by_i \mid \bX_{i\cdot}] \right|^r,
\]
where $\mathbb{E}[\by_i \mid \bX_{i\cdot}]$ is fitted by a non-linear outcome model $g(\cdot)$.
In the case of a linear model, where the gradient is constant across the input space, this reduces to:
$T_j = \left| \beta_j \right|^r$.

\subsection{MALD for Categorical Variables}

For one-hot encoded categorical variables, the local gradient corresponds to the difference in the model prediction between levels of the variable. For a binary variable $X_{ij} \in \{0,1\}$, the local derivative can be defined as:
\[
    \ell_{ij} = \mathbb{E}[\by_i \mid \bX_{ij}=1, \bX_{i,-j}, \wt{\bX}_{i\cdot}] - \mathbb{E}[\by_i \mid \bX_{ij}=0, \bX_{i,-j}, \wt{\bX}_{i\cdot}].
\]
For any multi-level categorical variable $j$ with levels $k=1,2,\ldots,K_j$, we define the local gradient score as:
\begin{gather*}
    \ell_{ij} =  \max_k \{g(X_{ij} = k, \bX_{i,-j},  \wt{\bX}_{i\cdot}) , 0\} - \min_k \{ g(X_{ij} = k, \bX_{i,-j}, \wt{\bX}_{i\cdot}) , 0\} \ge 0, \\
    \ell_j = \frac{1}{n} \sum_{i=1}^n \ell_{ij}^r.
\end{gather*}
Here, $g(\cdot)$ denotes the fitted outcome model, and $\{0\}$ is added to reflect the implicit coefficient zero of the baseline category in one-hot encoding.
We similarly define $\wt{\ell}_j$ for the knockoff feature $\wt{\bX}_{\cdot j}$: $\wt{\ell}_{ij} =  \max_k \{g(\wt{X}_{ij} = k, \wt{\bX}_{i,-j},  {\bX}_{i\cdot}) , 0\} - \min_k \{(\wt{X}_{ij} = k, \wt{\bX}_{i,-j},  {\bX}_{i\cdot}), 0\} \ge 0$, and $\wt{\ell}_j = \frac{1}{n} \sum_{i=1}^n \ell_{ij}^r.$
This formulation ensures that the importance measure for categorical features is both baseline-invariant and comparable in scale to that of continuous features. 

\paragraph{Implementation.} We implement MALD with two commonly used non-linear models: neural networks (NNs) and random forests. Specifically:
\begin{itemize}
    \item A key advantage of using MALD with NNs is that the gradients for all samples are computed and stored efficiently. Therefore ${\ell}_j$'s and $\wt{\ell}_j$'s can be obtained with little extra effert, i.e. almost for free when fitting NNs.
    \item Random forests as $g(\cdot)$, we use the following local approximation:
    \[
        \frac{\partial}{\partial X_{ij}} g(\bX_{i\cdot}) \approx \frac{g(\bX_{i\cdot} + b e_j) - g(\bX_{i\cdot})}{b},
    \]
    where $b > 0$ is a small bandwidth and $e_j$ is the $j$th standard basis vector.
\end{itemize}
Algorithm \ref{alg:local-grad-0} gives the {\bf MALD} implementation in detail. 
\begin{algorithm}
    \caption{Mean Absolute Local Derivative (MALD) Importance}
    \label{alg:local-grad-0}
    \begin{algorithmic}[1]
        \REQUIRE Predictive model $\wh{g}$ fitted on $(\by, [\bX, \wt{\bX}])$
        \INPUT Data $\left(\by, \bX, \wt{\bX} \right)$, bandwidth $b > 0$, exponent $r > 0$
        \OUTPUT Importance scores $W_j$ for $j = 1, \ldots, p$
        
        \FOR{$j = 1,\ldots, p$}
                \IF{$\bX_{\cdot j}$ is numeric}
                    \IF{exact derivative $\partial \wh{g} / \partial \bX_{ij}$ available}
                        \STATE $\ell_{ij} = \left| \frac{\partial \wh{g}(\bX_{i\cdot})}{\partial \bX_{ij}} \right|$;\quad $\wt{\ell}_{ij} = \left| \frac{\partial \wh{g}(\bX_{i\cdot})}{\partial \wt{\bX}_{ij}} \right|$
                    \ELSE
                        \STATE $\ell_{ij} = \left| \frac{\wh{g}(\bX_{i\cdot} + b e_j) - \wh{g}(\bX_{i\cdot})}{b} \right|$; \quad $\wt{\ell}_{ij} = \left| \frac{\wh{g}(\bX_{i\cdot} + b \wt{e}_j) - \wh{g}(\bX_{i\cdot})}{b} \right|$ 
                    \ENDIF
                \ELSIF{$\bX_{\cdot j}$ is categorical with levels $1,\ldots,K_j$}
                    \STATE Let $\mathcal{G}_{ij} = \left\{ \wh{g}(\bX_{i, -j}, \bX_{ij} = k) : k = 1, \ldots, K_j \right\}$
                    \STATE $\ell_{ij} = \left| \max \mathcal{G}_{ij} - \min \mathcal{G}_{ij} \right|$;
                    \State $\wt{\ell}_{ij} = \left| \max \wt{\mathcal{G}}_{ij} - \min \wt{\mathcal{G}}_{ij} \right|$ analogously
                \ENDIF
            \STATE $\ell_j = \frac{1}{n} \sum_{i=1}^n \ell_{ij}^r$
        \ENDFOR
        \STATE Return $W_j = \ell_j - \wt{\ell}_{ij}$
    \end{algorithmic}
\end{algorithm}

\section{Simulation Study}\label{sec-sim-0}

This simulation study has two main goals: (1) to compare different knockoff generation methods, and (2) to evaluate several importance measures, under varying data structures and outcome models.

The traditional second-order knockoff method performs well under uni-modal Gaussian distributions. Therefore, our primary focus is on more complex settings with heterogeneous and multimodal covariates. 
In total, we compare a linear outcome using cross-validated LASSO importance against three nonlinear outcome models evaluated using nonlinear importance measures.

\paragraph{Covariate Generation}
We simulate covariates \( X \in \mathbb{R}^{n \times p} \) as a mixture of Gaussian distributions, where each mode has the same covariance structure but a distinct mean. We consider three settings, each varying one of the following: signal magnitude, number of mixture modes, or intra-mode correlation. The simulation parameters are summarized in Table~\ref{tab:sim-settings}.

\begin{table}[h]
\centering
\begin{tabular}{l|c|c|c}
\textbf{Label} & Signal strength & \# of modes $m$ & AR(1) Correlation $\rho$  \\
\hline
Signal Varying       & $1,2,4,8,16,32$ & 5   & 0.50  \\
Modes Varying        & 8              & $1,3,5$ & 0.50  \\
Correlation Varying  & 8              & 5   & $0.00, 0.25, 0.50, 0.75$  \\
\end{tabular}
\caption{Simulation settings for covariate generation.}
\label{tab:sim-settings}
\end{table}
We fix sample size $n=1024$ and $p = 128$ covariates: 96 numeric and 32 categorical. Among categorical variables, 16 have two levels, and 16 have three levels. 

\subsection{Outcome Models}\label{section:simulation-outcome-0}

Each covariate configuration is paired with two outcome functions—one linear and one nonlinear—using the same active covariates. For simplicity, we denote \( X_j \) as the $j$-th covariate and \( I(\cdot) \) as an indicator function.

\paragraph{Linear Outcome}
The linear outcome is given by:
\begin{equation}\label{eq:linear-y}
\begin{aligned}
y =\;& \beta(X_{2} - X_{7} + X_{31} - X_{86} + X_{87}) \\
&+ \beta(-2\cdot I(X_{98} = 1) + 2\cdot I(X_{98} = 2)) \\
&+ \beta(-2\cdot I(X_{99} = 1) + 2\cdot I(X_{99} = 2)) \\
&+ \beta(1\cdot I(X_{113} = 1) - 2\cdot I(X_{113} = 2) - 2\cdot I(X_{113} = 3)) \\
&+ \beta(-2\cdot I(X_{126} = 1) - 1\cdot I(X_{126} = 2) + 1\cdot I(X_{126} = 3)) \\
&+ \beta(2\cdot I(X_{128} = 1) - 2\cdot I(X_{128} = 2) + 1\cdot I(X_{128} = 3)) \\
&+ \epsilon
\end{aligned}
\end{equation}

\paragraph{Nonlinear Outcome}
The nonlinear outcome uses the same covariates but applies nonlinear transformations to numeric features:
\begin{equation}\label{eq:nonlinear-y}
\begin{aligned}
y =\;& \beta \cdot 3.76 \cdot \frac{1}{1 + X_{2}^2} 
- \beta \cdot 1.94 \cdot \log(1 + X_{7}^2) 
+ \beta \cdot 1.42 \cdot \sin(2\pi X_{31}) \\
&+ \beta \cdot (X_{86} X_{87} - 0.25 X_{86} + 0.25 X_{87}) \\
&+ \text{(same  terms for categorical variables as in Eq.~\ref{eq:linear-y})} 
+ \epsilon
\end{aligned}
\end{equation}
The numerical coefficients (e.g., 3.76, 1.94) are chosen so that each nonlinear term has unit variance when evaluated on a standard normal input, aligning effect magnitudes across transformations. The error term is independently sampled as $\epsilon_i \sim \mathcal{N}(0,1)$.

\subsection{Knockoff Methods}

The following knockoff methods were implemented for comparison, where CF-Second and CF-GAN are the proposed ones based on conditional residuals and using random forests to fit the conditional expectations.
\begin{enumerate}
    \item GAN: Generative Aversarial Network
    \item Second: Second Order (Gaussian) Knockoffs
    \item CF-Second: Conditional Forests with Second Order Residuals
    \item CF-GAN: Conditional Forests with GAN Residuals
    \item Forest Sequential Conditional Independent Pairs (SCIP) with Permuted Residuals
\end{enumerate}

GAN Knockoffs are one example of fully nonparametric knockoffs from an adversarial neural network which minimizes the ability to distinguish between the distribution of $X$ and $\wt{X}$ while penalizing high pairwise correlation between each $X_j$, $\wt{X}_j$. It was not specifically designed for categorical variables, but can handle them via one-hot encoding.
Second Order Knockoffs are counterfactual except in the case of just one mode; it generates knockoffs as if $X$ came from a unimodal Gaussian distribution, while minimizing the pairwise correlation between each $X_j$, $\wt{X}_j$. It serves as another baseline comparison.

The two conditional forest methods use algorithm \ref{alg:conditional-residual-estimator-heterogeneous-0}. For the knockoff residual portion, they use GAN with the residuals for CF GAN (Conditional Forest GAN), and regular second order knockoffs for CF Gaussian.
The Forest Sequential Conditional Independent Pairs (Forest SCIP) method uses a random forest for each numeric variable, and then permutes the residuals among all data points for each variable, described in algorithm \ref{alg:SCIF-estimator-heterogeneous-0}.
 \emph{R} package  \verb|ranger| was used to fit random forests, with the importance measure and conditional residuals and implemented with \verb|rangerKnockoff|.

\begin{algorithm}
    \caption{Sequential Conditionally Independent Forests for Heterogeneous Data
    \label{alg:SCIF-estimator-heterogeneous-0}}
    \begin{algorithmic}[1]
        \REQUIRE Predictive models $\Psi_j$ for each variable $j \in \{1,\ldots,p\}$, knockoff generator $C$
        \INPUT Original data matrix $\bX \in \mathbb{R}^{n \times p}$
        \OUTPUT Knockoff matrix $\wt{\bX} \in \mathbb{R}^{n \times p}$

        \FOR{$j = 1, \ldots, p$}
            \STATE Fit model $\wh{g}_j = \Psi_j(\bX_{\cdot j}, [\bX_{\cdot,-j}, \wt{\bX}_{\cdot,1:(j-1)}])$
            \IF{$\bX_{\cdot j}$ is numeric}
                \STATE Compute predictions $\wh{\bX}_{\cdot j} = \wh{g}_j([\bX_{\cdot,-j}, \wt{\bX}_{\cdot,1:(j-1)}])$
                \STATE Compute residuals $\wh{\bGamma}_{\cdot j} = \bX_{\cdot j} - \wh{\bX}_{\cdot j}$
                \STATE Generate knockoff residuals $\wt{\bGamma}_{\cdot j} = C(\wh{\bGamma}_{\cdot j})$
                \STATE Set $\wt{\bX}_{\cdot j} = \wh{\bX}_{\cdot j} + \wt{\bGamma}_{\cdot j}$
            \ELSIF{$\bX_{\cdot j}$ is categorical}
                \STATE Estimate category probabilities $P(\bX_{ij} = k) = \wh{g}_j(k; \left[\bX_{i,-j}, \wt{\bX}_{i,1:(j-1)} \right])$
                \STATE Sample $\wt{\bX}_{ij}$ from the estimated distribution
            \ENDIF
        \ENDFOR
        \STATE Return $\wt{\bX}$
    \end{algorithmic}
\end{algorithm}

\subsection{Importance Measures}

The linear importance measure uses the cross validated LASSO with one-hot encoded categories, implemented in Python's \verb|sklearn.linear_model.LassoCV|.
In addtion, we compare three different nonlinear importance measures.
\begin{enumerate}
	\item The Gini coefficient is the baseline, previous best importance measure for nonlinear outcomes with random forests. It is implemented in \emph{R} \verb|knockoff| via \verb|::stat.random_forest| with default settings after one-hot encoding categorical variables.
	
	\item The MALD measure with random forests is implemented in \emph{R} \verb|rangerKnockoff|, using default \verb|ranger| settings to fit a random forest, and estimates the local derivatives with bandwidth $n^{-1/5}$.
	
	\item The MALD measure with neural networks is implemented in Python \verb|maldimportance| by training a tensorflow 2 layer dense neural network and using auto differentiation. 
	
\end{enumerate}

\subsection{Simulation Results}

\begin{figure}
    \centering
    \caption{Simulation results with scenarios of varying effect sizes.}
    \includegraphics[width=0.85\linewidth]{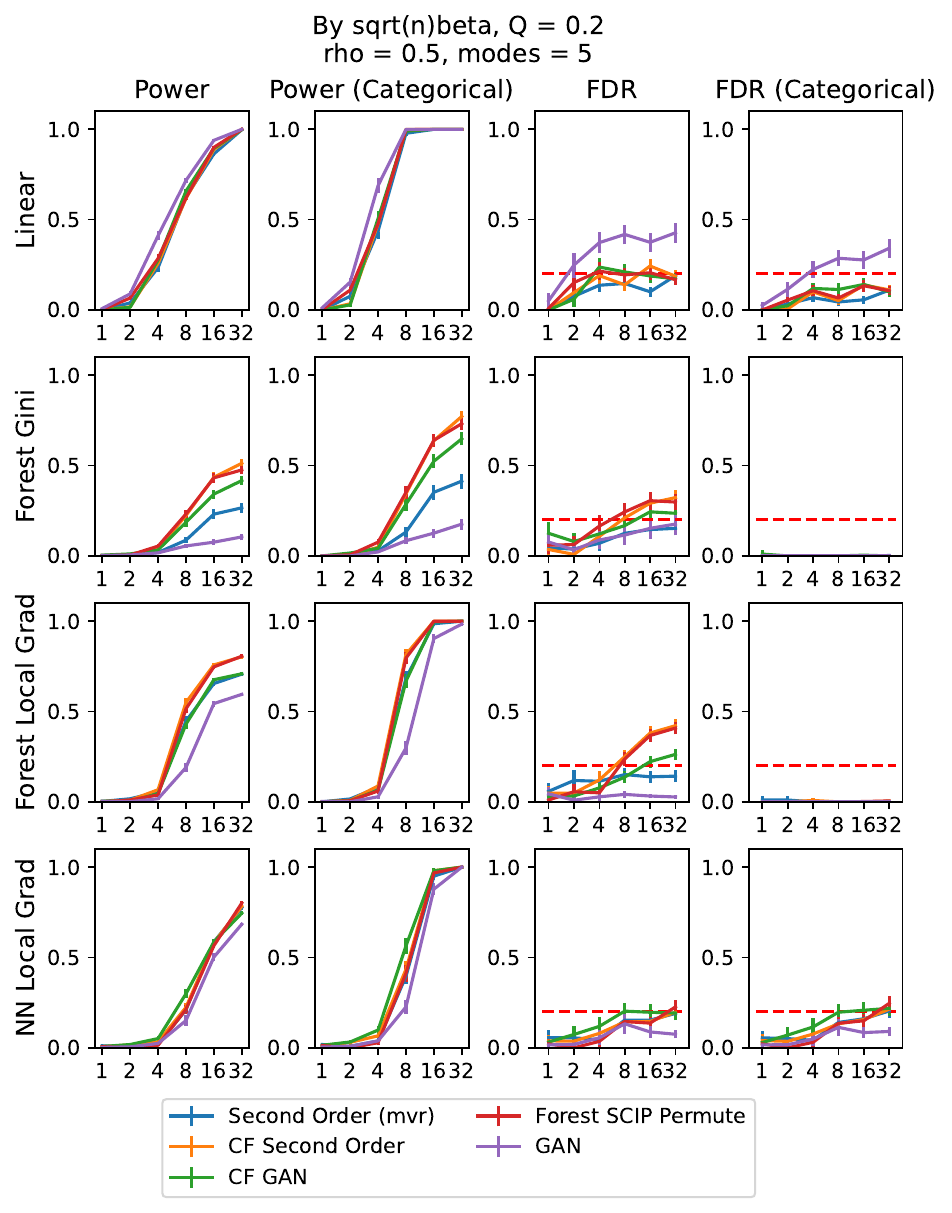}
    \label{fig:sim-beta-0}
\end{figure}

\begin{figure}
    \centering
    \caption{Simulation results with scenarios of varying covariate modes.}
    \includegraphics[width=0.85\linewidth]{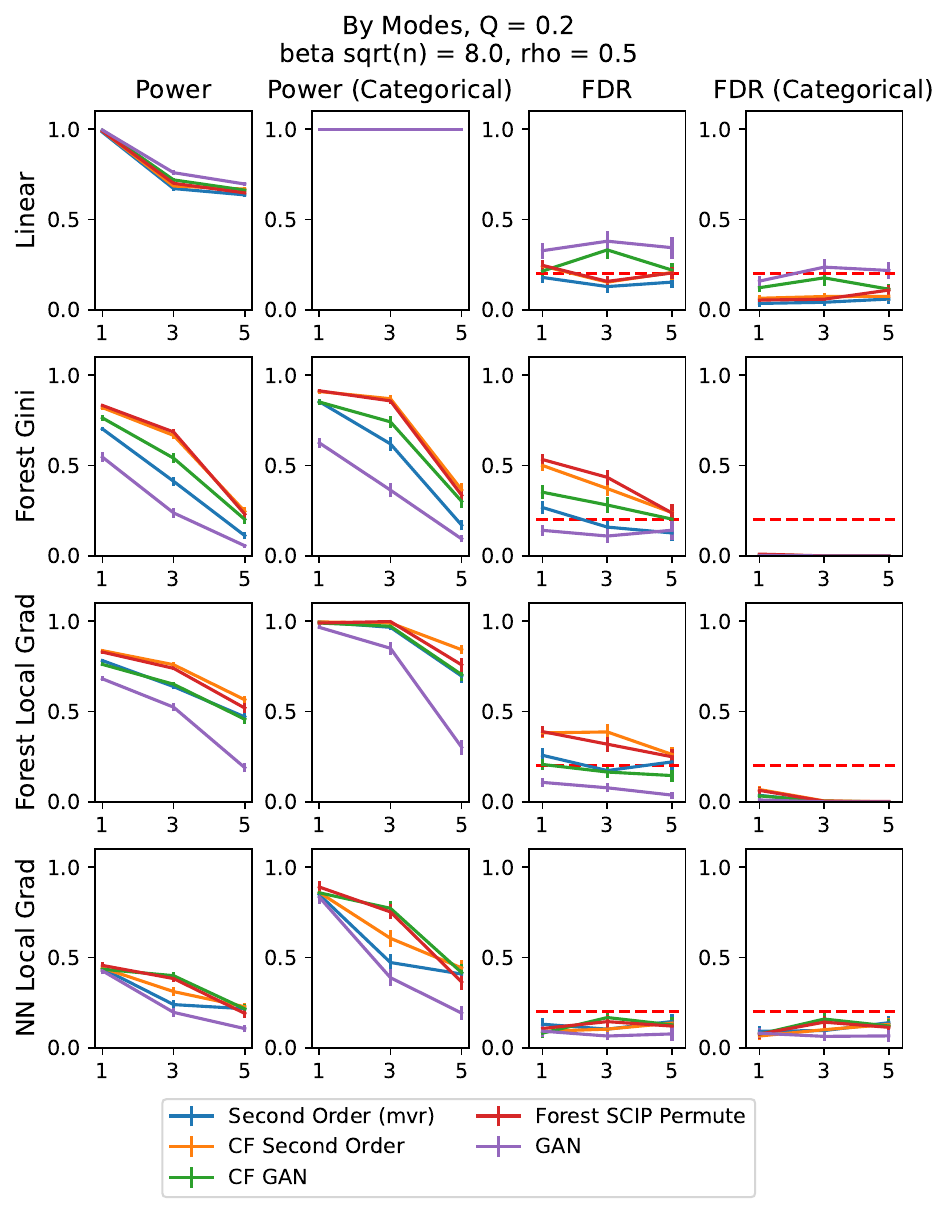}
    \label{fig:sim-modes-0}
\end{figure}

\begin{figure}
    \centering
    \caption{Simulation results with scenarios of varying covariate correlations.}
    \includegraphics[width=0.85\linewidth]{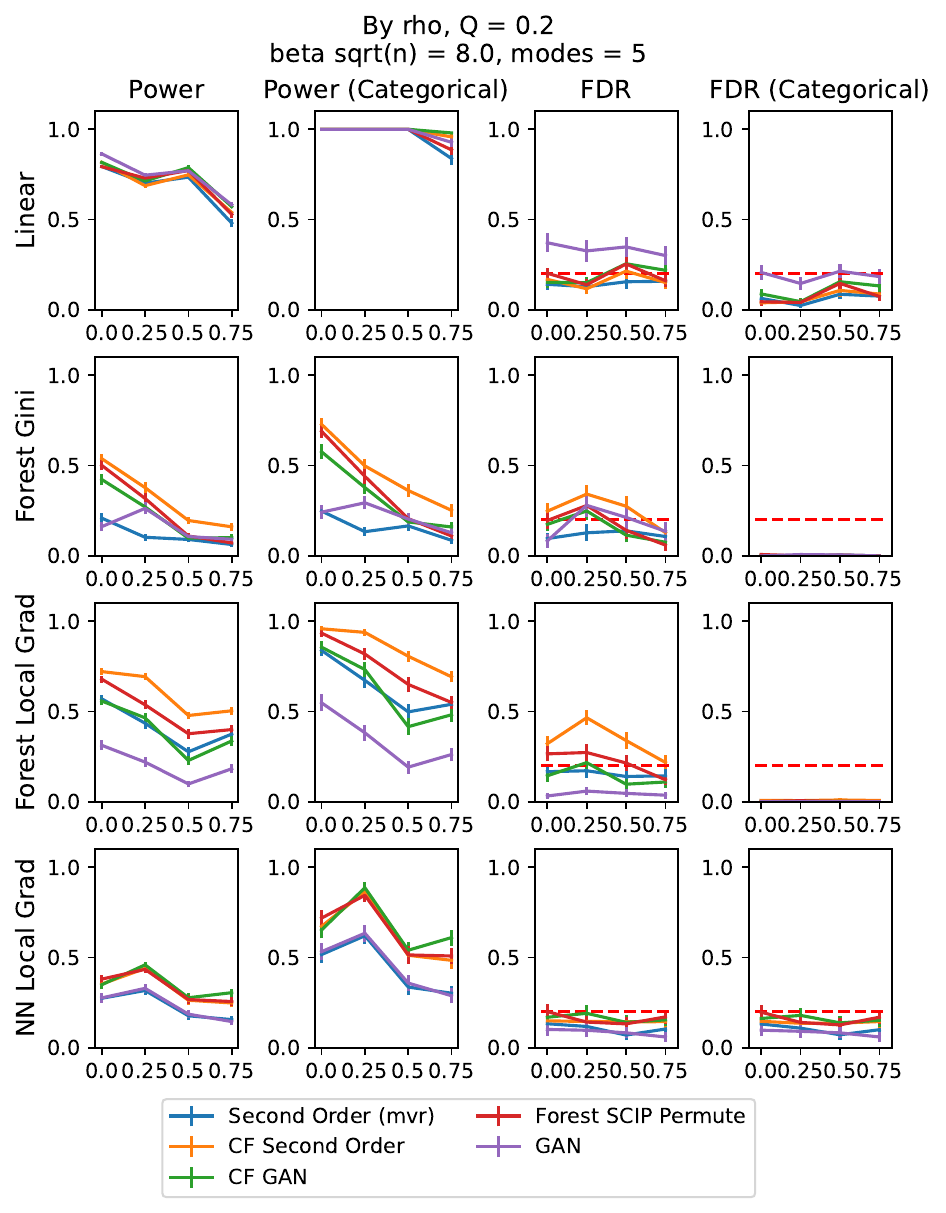}
    \label{fig:sim-rho-0}
\end{figure}

Figures~\ref{fig:sim-beta-0}, \ref{fig:sim-modes-0}, and \ref{fig:sim-rho-0} summarize the power and false discovery rate (FDR) of all knockoff generation and importance measure combinations across different scenarios.

Overall, the Second-order knockoffs perform very well in settings where the outcome is linear, exhibiting high power and accurate FDR control. However, when applied to nonlinear outcomes, they tend to be overly conservative, yielding low FDR but reduced power. This conservativeness diminishes somewhat in unimodal covariate distributions but remains substantial in heterogeneous mixtures.

The Generative Adversarial Network (GAN)-based knockoffs show somewhat opposite patterns. In linear outcome scenarios, GAN knockoffs often produce too many selections, leading to inflated FDRs above the nominal level. Conversely, with nonlinear outcomes, they become highly conservative, resulting in low FDR but also noticeably diminished power.

Among the 3 proposed methods, including Conditional Residual Knockoffs (CF-Second, CF-GAN) and the Forest-based SCIP variant (Forest-SCIP), performance is generally strong for linear outcomes, closely matching the Second-order knockoffs in both FDR and power. For nonlinear outcomes, both the CF-Second and Forest-SCIP have the highest powers, while CF-GAN has the best FDR control with comparable power to the other two. This empirical observation aligns with the theoretical motivation for conditional residual knockoffs.

Regarding importance measures, the Random Forest-based measures—both the change in Gini impurity and the Local Gradient (MALD) estimates—tend to be less conservative. They deliver higher power but sometimes at the cost of exceeding the nominal FDR threshold, especially when used with Forest SCIP Permute or CF Second-Order knockoffs. The neural network-based MALD (NN MALD) shows more stable FDR control while achieving power levels that are often nearly as high as the other two importance measures.

In summary, these results indicate that the Conditional Forest GAN method is the most robust and consistent knockoff generation strategy for heterogeneous data with nonlinear outcome models. Among importance measures, the Forest-based MALD approach offers an effective combination of power and reliable FDR control.

\section{Real Data Analysis}\label{section:real-data-0}
\begin{figure}[ht]
    \includegraphics[width=0.45\textwidth]{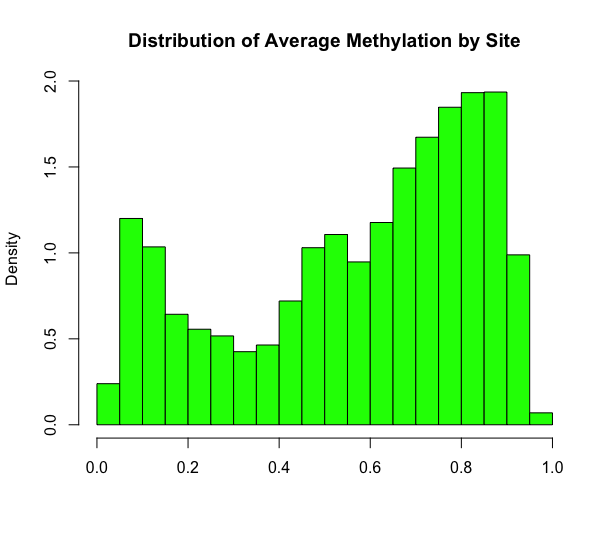}
    \includegraphics[width=0.45\textwidth]{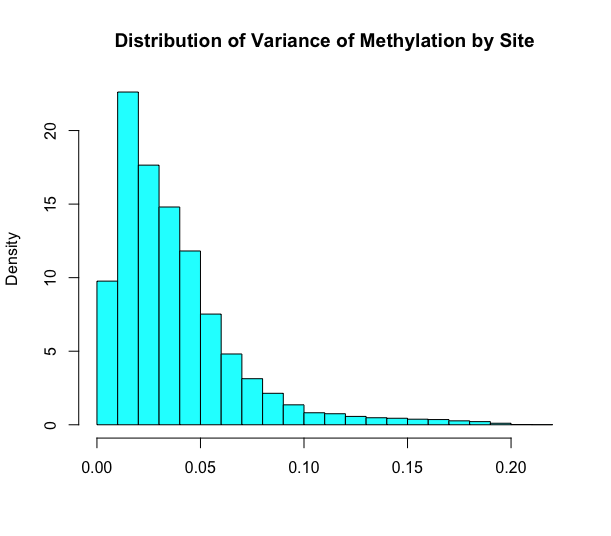}
    \caption{Mouse methylation data distributions.}
    \label{fig:methylation-0}
\end{figure}

The Mammalian Methylation Consortium has assembled an extensive dataset comprising over 15,000 individual samples from more than 180 mammalian species and 59 tissue types \citep{lu2023universal,haghani2023dna}. A novel mammalian methylation array was developed \citep{arneson2022mammalian}, and
DNA methylation measurements were collected at 37,554 CpG sites, with values ranging continuously between 0 and 1. Numerous aging clocks have been developed using subsets of these data to predict chronological age; further details and resources are available at \url{https://github.com/shorvath/MammalianMethylationConsortium}. The principal goal of these models is to predict an individual's chronological age, a continuous numeric outcome. A standard preprocessing step involves reducing the dimensionality of the predictors (CpG sites) by unsupervised screening. Figure~\ref{fig:methylation-0} shows the distribution of the means and variances for methylation at each CpG site.

While there have been numerous studies developing DNA methylation clocks specifically for mouse data, these approaches typically assume a linear relationship between age and methylation levels and may fail to capture the heterogeneity of effects across different tissue types. To address this limitation, we analyze a subset of 5,562 mouse samples—hereafter referred to as the \emph{Mouse Data}—using our proposed nonparametric knockoff methods. The objectives are twofold: to validate previously reported findings on this dataset and to identify truly important CpG sites while rigorously controlling the false discovery rate (FDR).

As a reference for potentially relevant CpG sites, we consider the analysis presented by \citet{murine_genetic_loci}, who fit a series of LASSO models across multiple species, including the mouse data. The outcome variables from that study include several measures of biological development and age and are available on GitHub at \url{https://github.com/jazoller96/mammalian-methyl-clocks}.
Specifically, we focus on the following sets of LASSO-derived coefficients:
\begin{itemize}
    \item Developmental Clocks: \emph{Coef.MousePanTissueDevelopmentalClock}
    \item Interventional Clocks: \emph{Coef.MousePanTissueInterventionalClock}
    \item Age Clocks: \emph{Coef.MousePanTissueClock} and \emph{Coef.MouseLiverClock}
\end{itemize}
Our primary interest lies in the latter two. The Pan-Tissue Clock identifies CpG sites predictive of age across all tissue types, whereas the Liver Clock highlights sites predictive of age specifically within liver tissue, the most frequently sampled tissue in the dataset. Overall, 537 CpG sites were identified as significant in at least one of these mouse age clocks.

Table~\ref{tab:tissue-counts-0} summarizes the distribution of tissue types among the Mouse Data samples. Additional details are provided in Appendix~\ref{section:appendix-tissue-0}.

\begin{table}[ht]
    \centering
    \begin{tabular}{l|r}
        \textbf{Tissue Type} & \textbf{Sample Count} \\
        \hline
        Liver      & 1,436 \\
        Blood      & 865 \\
        Cortex     & 612 \\
        Muscle     & 428 \\
        Striatum   & 349 \\
        Cerebellum & 264 \\
        Heart      & 226 \\
        Kidney     & 179 \\
        Skin       & 142 \\
        Other      & 1,061
    \end{tabular}
    \caption{Distribution of tissue types among 5,562 mouse samples.}
    \label{tab:tissue-counts-0}
\end{table}

In our analysis, we include all 537 CpG sites previously identified as potentially relevant, along with mouse sex and tissue type as categorical covariates. In principle, CpG sites from both the Pan-Tissue and Liver Clocks should emerge as significant since our data include all tissue types. Additionally, our nonlinear models should, at least asymptotically, be able to detect sites whose effects are tissue-specific. However, given the finite sample size, some loss of power is expected compared to the original studies.

Among the remaining CpG sites not flagged as significant in \citet{murine_genetic_loci}, we select the 200 sites with the highest marginal variance, reasoning that higher-variance sites may contain more informative variation. This yields a final predictor set of 737 methylation sites for knockoff generation, modeling, and selection. Because nonparametric methods are relatively insensitive to finer distinctions among tissue types, we further aggregate tissue annotations into seven broader categories, as detailed in Appendix~\ref{section:appendix-tissue-0}.

\subsection{Synthetic Mouse Data}\label{section:simulation-mouse-0}

First, using the real CpG/Tissue/Age mouse data and the corresponding Knockoffs, we generate fully synthetic outcome data as a linear function of the real dataset. Each of 5 different outcome sets has a different random selection of  23 of the numeric explanatory variables, and both the categorical variables, sex and tissue. For each selected variable, after normalizing to mean 0, variance 1, we multiply the value by $\pm \beta_j/\sqrt{n}$, positive or negative with probability $0.5$.

With the number of variables $p$ and datapoints $n$ in the mouse data, the nonparametric outcome methods have difficulties with power. More realistically, we are more likely to find effects which are strong and with some monotonic effect. The LASSO is good for these cases, and matches the method used on the transformed age in the original analysis. We test two values of signal strength $|\bbeta_j|$, summarized in table \ref{tab:synthetic-0}.

\begin{table}[]
    \centering
    \begin{tabular}{l | l | l | l}
         Label & $|\beta_j|\sqrt{n}$ Numeric & $|\beta_j|\sqrt{n}$ Categorical & Relevant Variables  \\
         \hline
         Weak Signal & $12$ & $\{ 6, 12 \}$ & 25\\
         Strong Signal & $18$ & $\{ 9, 18 \}$ & 25
    \end{tabular}
    \caption{Synthetic outcome generation.}
    \label{tab:synthetic-0}
\end{table}

The results with the nonlinear outcome and corresponding measure use the same explanatory data, but instead choose a random selection of 23 nonparametric functions, and coefficients for both categorical variables. The choice of nonlinear functions include those with the simulations from section \ref{section:simulation-outcome-0} and more:
\begin{itemize}
    \item Cauchy: $3.76\frac{1}{1 + \bX_{ij}^2}$
    \item Log: $1.94\log(1 + \bX_{ij}^2)$
    \item Square: $0.7 \bX_{ij}^2$
    \item Sin: $1.42 \sin(2\pi\bX_{ij})$
    \item Cos: $1.42 \cos( 2\pi\bX_{ij})$
    \item Square Root: $2.86 \sqrt{|\bX_{ij}|}$
    \item Linear: $\bX_{ij}$
    \item First order interaction: $\bX_{ij}\bX_{ik} \pm 0.25\bX_{ij} \pm 0.25\bX_{ik}$
\end{itemize}

The goal is to see if the newly proposed methods are competitive in picking out known variables, even when existing methods should be adequate for a linear model.

\begin{figure}[ht]
    \centering
    \includegraphics[width=0.7\linewidth]{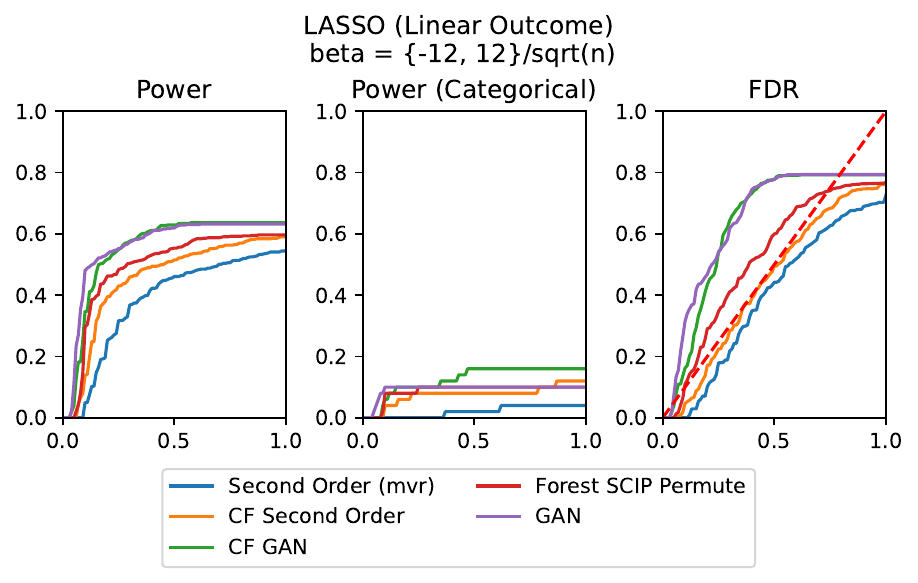}
    \includegraphics[width=0.7\linewidth]{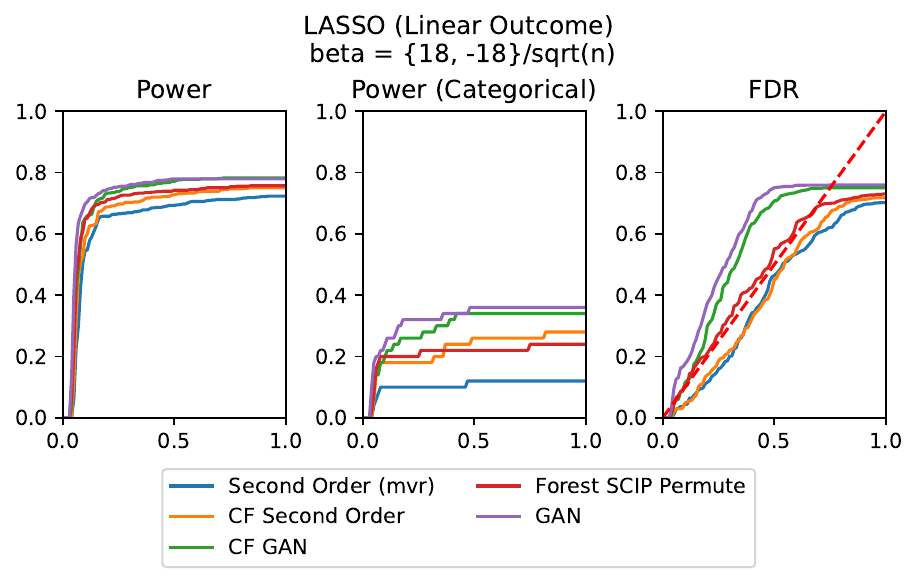}
    \caption{Synthetic linear outcomes with mouse data.}\label{fig:mouse-linear-y-0}
\end{figure}

The two cases, demonstrated in Figure \ref{fig:mouse-linear-y-0} show that the GAN and CF GAN methods consistently violate the FDR constraint. This shows that they do not in fact make appropriate knockoffs, so they lose the effect of acting as a negative control, since they are too different from the real $\bX$ which are irrelevant, but perhaps highly correlated with relative variables. The Conditional Forest with Second Order Residuals, Forest SCIP, and Second Order methods all have consistent FDR control. However, the Conditional Forest with Second Order residuals has better power while still being below the nominal fdr target. This shows that with the larger number of variables, the methods using GAN are ineffective, but the two other nonparametric methods, both new, appear to have no downsides compared to Second Order Knockoffs.

\begin{figure}[ht]
    \centering
    \includegraphics[width=0.7\linewidth]{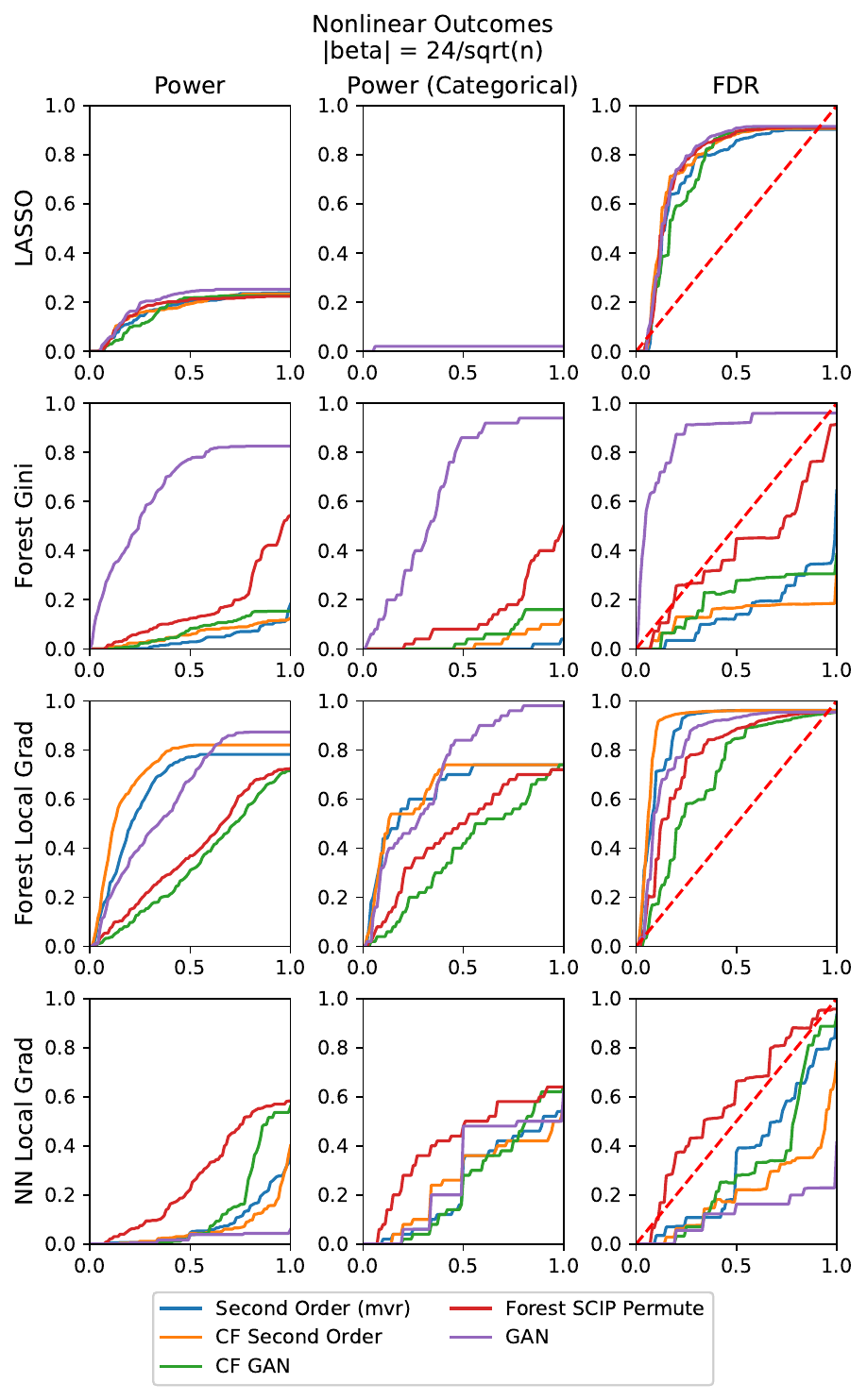}
    \caption{Synthetic nonlinear outcomes with mouse data, small effect size.}\label{fig:mouse-nonlinear-y-0}
\end{figure}

With a nonlinear outcome, however, the traditional methods of knockoffs ( Second Order and GAN) and the traditional outcome importance of the LASSO completely fail both any reasonable power, and total FDR control violation. Figure \ref{fig:mouse-nonlinear-y-0} shows the Target FDR to Power and True FDR curves with synthetic nonlinear outcomes. This is consistent with the expectation that nonlinear, heterogeneous estimators are difficult with data in around the size of $n\times p = 5562\times 739$. The results show significant improvement with new methods.

\begin{figure}[ht]
    \centering
    \includegraphics[width=0.7\linewidth]{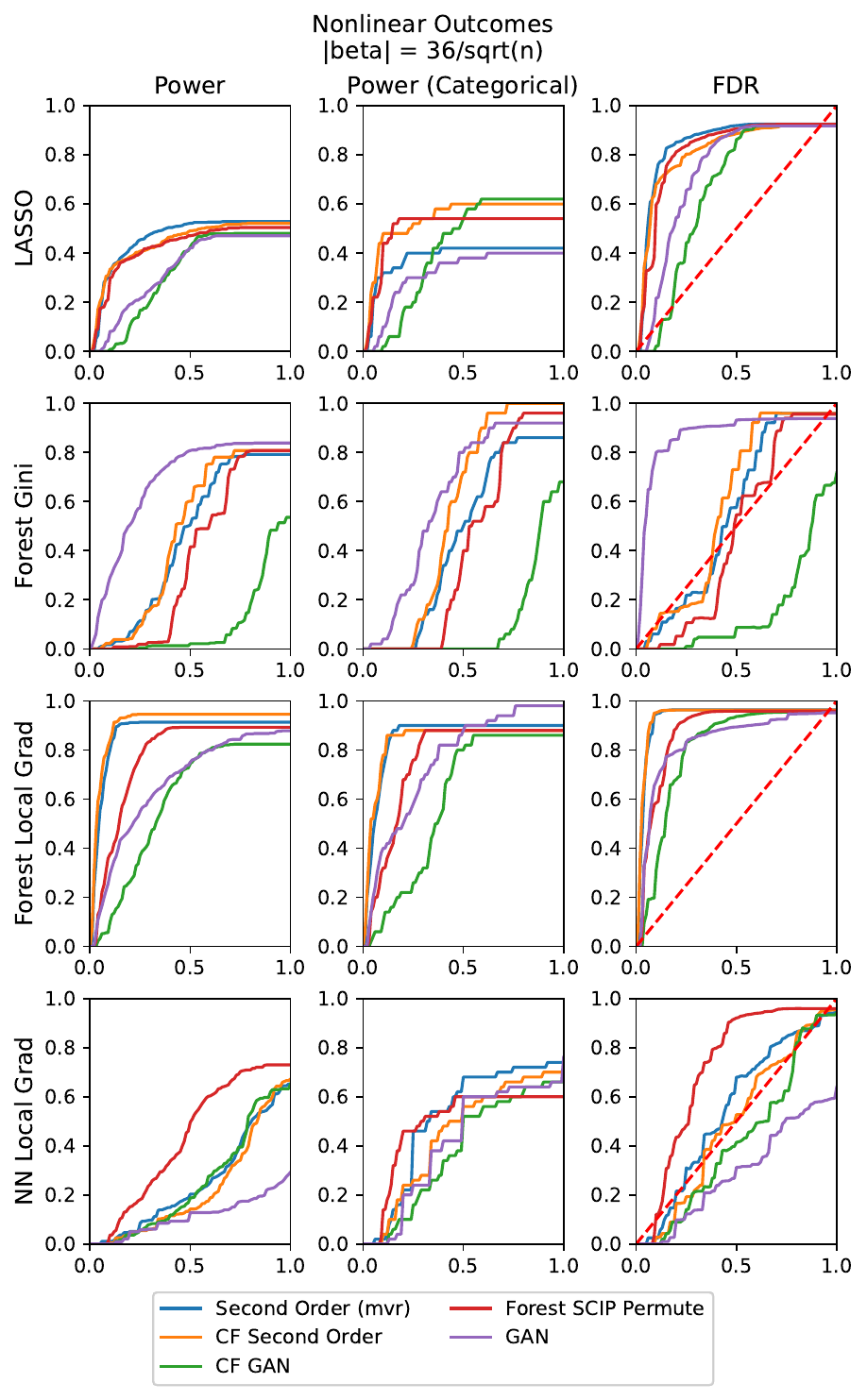}
    \caption{Synethetic nonlinear outcomes with mouse data, moderate effect size.}\label{fig:mouse-nonlinear-y-1}
\end{figure}

The forest methods do suffer weak power, but using the Gini measure, the default for random forests, does give better FDR control than the forest local gradient. Ultimately, however, the Neural Network Local Gradient appears to keep the most reliable FDR control, and the methods of Conditional Forest are among the most competitive. Figure \ref{fig:mouse-nonlinear-y-1} shows the more pronounced effect and rising power with a larger signal strength of $|\beta_j|$'s.

\begin{figure}[ht]
    \centering
    \includegraphics[width=0.45\linewidth]{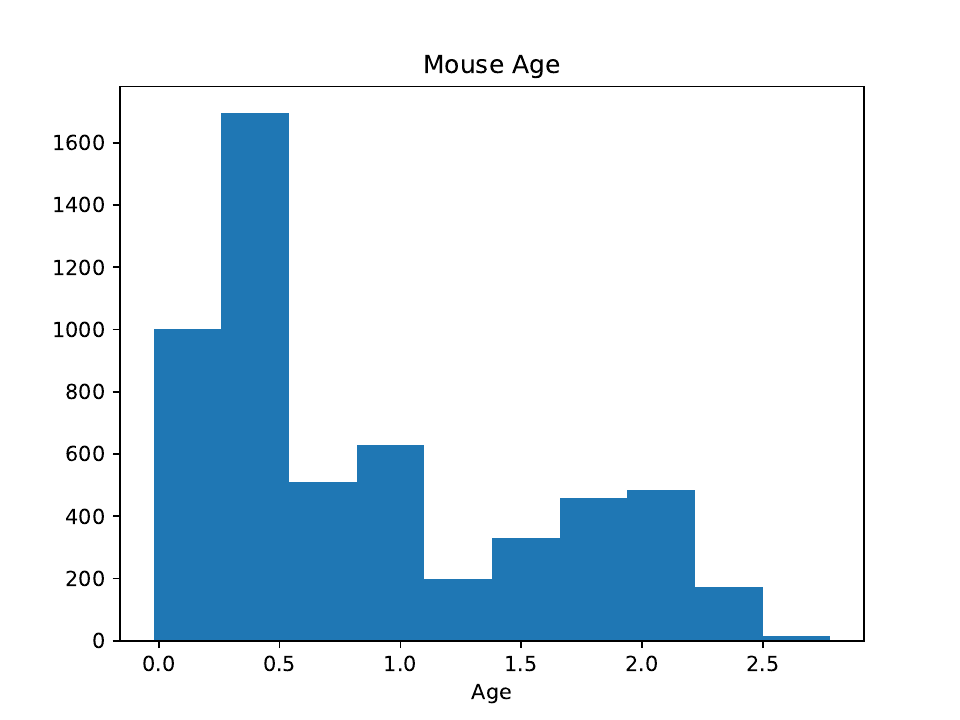}
    \includegraphics[width=0.45\linewidth]{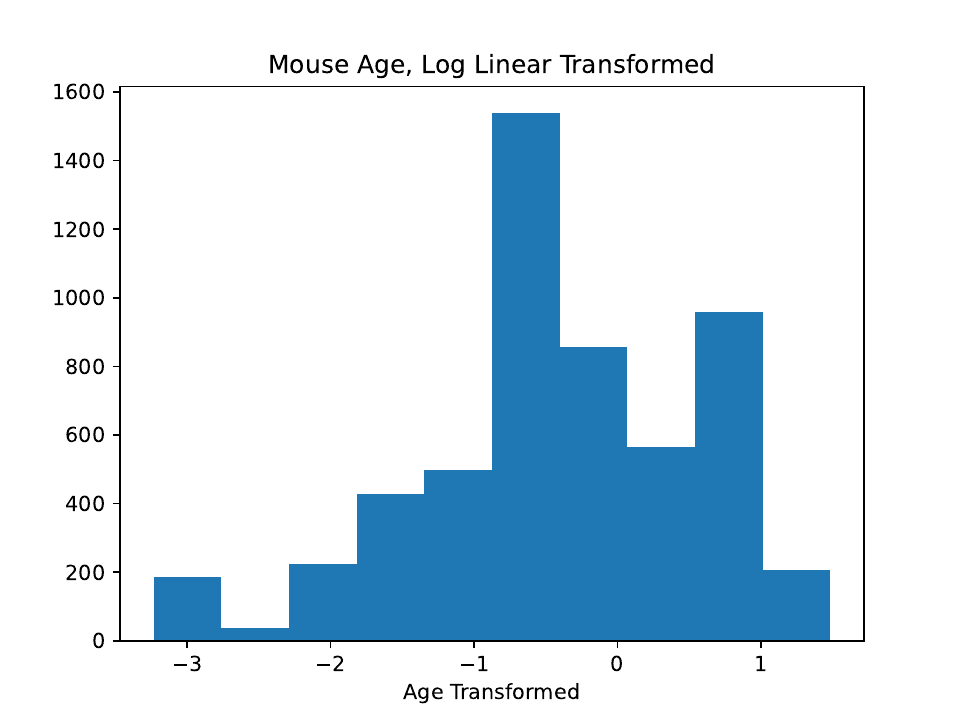}
    \caption{Mouse age distributions.}
    \label{fig:mouse-ages-0}
\end{figure}

\subsection{CpG Site Selection for Aging}

The results in \cite{lu2023universal} describe the variables selected with a LASSO for predicting the real transformed mouse age. We do the same, adding tissue type and sex as categorical predictors. We apply the same log-linear age transformation as in \cite{lu2023universal}, which can also be found in \url{https://github.com/jazoller96/mammalian-methyl-clocks/AgeTransformations/1-age_transformations_functions_Joseph_Zoller.R}  with the function in \verb||: \verb|fun_llinmouse.inv|:
$$
    f(\text{Age})= \begin{cases}
        \log( \text{Age} + 0.06), & \text{ if } \text{Age} < 1.2;\\
         (\text{Age} - 1.2)/(1.26) + \log(1.26), & \text{ otherwise}.
    \end{cases}
$$
Figure \ref{fig:mouse-ages-0} shows the distributions of raw age, and transformed age, illustrating that the transformed age is more normally distributed. 

\begin{figure}[ht]
    \centering
    \includegraphics[width=0.5\linewidth]{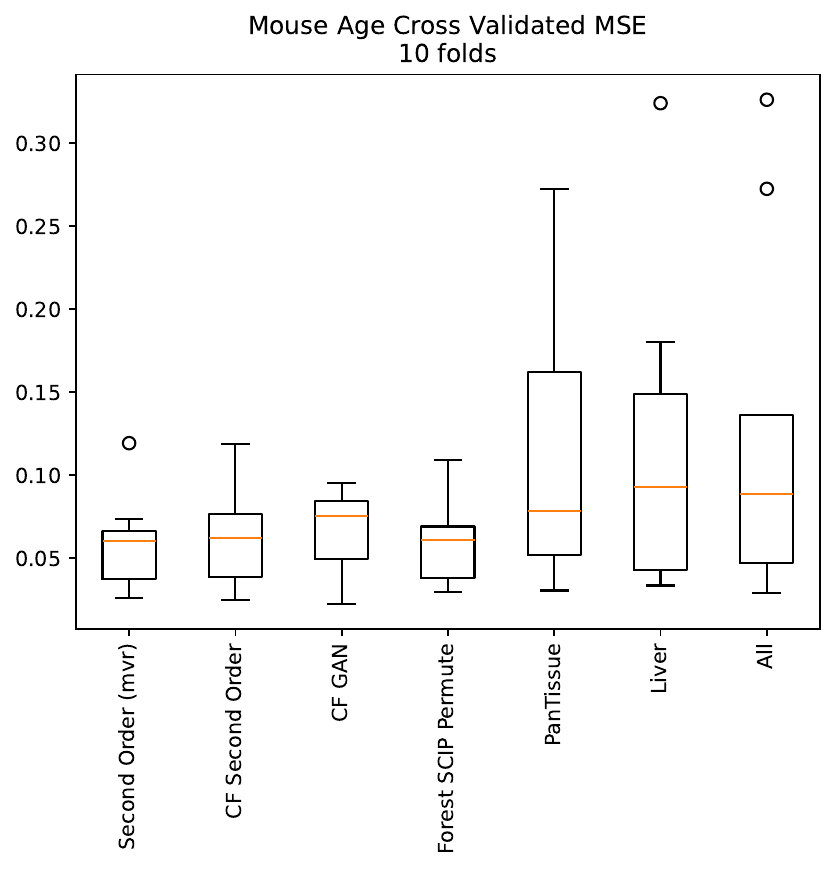}
    \caption{Cross validated errors by different models with transformed mouse age.}
    \label{fig:mouse-age-crossValidation-0}
\end{figure}

Table \ref{tab:crossValidation-summary-0} contains the cross validated mean squared error (MSE) results for the selection from different sets of selected variables; Figure \ref{fig:mouse-age-crossValidation-0} has box plot displays.
We only use the smallest number of variables among each of the 5 trials of each knockoff method, selected from the LASSO knockoff step to perform the cross validation.

The cross validation procedure uses 10 fold cross validation. With the selected set of variables only, we fit an ordinary least squares model with those variables from the training set. Then, with the fit model from the $9/10$ training data points, we calculate the mean squared error on all the remaining $1/10$ test points. Record this error; it is one of the data points in the box plots in table \ref{tab:crossValidation-summary-0}. Then, repeat this process, with a different test set and training set, leading to slightly different models and slightly different mean squared errors, using each data point exactly once as a test point, and nine times as a training point.
\begin{table}[ht]
    \centering
    \begin{tabular}{c | r | l | l}
         Selection Source & Variables Selected &  MSE & \\
             \hline
         All & 739 & 0.1219\\
         Pan Tissue & 537 & 0.1109\\
         Liver & 385 & 0.1147\\
         \hline
         Second Order & 184 & 0.0578\\
         CF Second Order & 177 & 0.0618\\
         CF GAN & 112 & 0.0664\\
         Forest SCIP Permute & 176 & 0.0593
    \end{tabular}
    \caption{Cross validated errors by different methods and selection results.}
    \label{tab:crossValidation-summary-0}
\end{table}
The results with each knockoff method clearly show a narrower range of mean squared errors with generally lower values than from the Pan Tissue, Liver, and All variable groups. The results show that every knockoff procedure with the exception of GAN performs extremely well compared to the baseline selections. This is in fact rather promising for the nonparametric knockoff methods: as we expect, the Second Order Method, the simplest of all knockoff methods, performs essentially as well as the others. This makes sense since we have a linear outcome model. Though the first and second moments are not a sufficient statistic for a cross validated LASSO, they are for a LASSO with fixed penalty $\lambda$ and for ordinary least squares. The examples in this paper consistently show that the second order method works well with all linear outcomes. But the Conditional Forest (CF) Second Order methods and Forest SCIP Permute methods for constructing knockoffs perform no worse, perhaps even better, despite the fact that these nonlinear, nonparametric methods can have considerably more fitting noise with data sets which are not especially long; typically they work better with a larger $n/p$ ratio. This suggests that in real applications, they may be safe choices to use whether or not a linear model turns out to be correct.

The knockoff methods using the smaller set of variables resulting in lower cross validated error suggests that the original work on the developmental clocks did make many false discoveries for which the knockoff procedure can successfully control.

\section{Conclusions}

We proposed flexible, nonparametric methods for knockoff generation and variable importance, relaxing the typical Gaussian and linear model assumptions. Our aim was to retain statistical power and FDR control while improving adaptability to heterogeneous and nonlinear data structures.

We focused on two knockoff generation strategies: (1) Conditional Residual (CR) knockoffs using random forests, and (2) Sequential Conditionally Independent Pairs (SCIP) also implemented via forests. Combined with novel or existing importance measures, these knockoffs lead to variable selection through the knockoff filter, using scores of the form $W_j = \ell_j - \wt{\ell}_j$ and controlling FDR at nominal levels (e.g., 0.2).

To handle nonlinear outcome models, we introduced the Mean Absolute Local Derivative (MALD) importance, which outperforms LASSO-based measures when the outcome model cannot be captured linearly. Simulations and empirical results show that our forest-based knockoff methods consistently outperform both traditional second-order knockoffs and neural network alternatives in terms of power, FDR control, and interpretability—particularly under mixed, non-Gaussian covariates and nonlinear outcome models.

Despite these strengths, our approach entails several limitations:
Our CR knockoffs for categorical variables rely on a simplifying assumption that knockoff variables are conditionally independent given the non-categorical features:
\[
\{j, k \in J_C\}, j \neq k \ \Rightarrow \ \wt{\bX}_{ij} \perp \wt{\bX}_{ik} \mid \bX_{i,-\{j,k\}}.
\]
This allows marginal prediction via $g_j(\bX_{i,-j})$ to approximate categorical knockoff distributions. Although this ignores the full joint dependency among knockoff categories, simulations (with $n = 1024$ to $5562$, $p = 128$ to $739$) show excellent empirical performance. Modeling the full joint distribution across categorical variables would require sample sizes exponential in the number of levels, $O(\prod_{j \in J_C} K_j)$.
Our methods work well at moderate to high dimensions. However, as $p$ increases much larger than $n$, power and FDR control may deteriorate and screening steps are required as illustrated in the real data application. Conversely, as $n$ increases, neural-network-based approaches (e.g., GAN, Deep Knockoffs, Autoencoders) may become competitive following the conditional residual derivation via random forests.

Overall, Conditional Residual Knockoffs with random forests represent a robust and interpretable default under complex, heterogeneous data. MALD provides a powerful, flexible importance measure when outcome models are nonlinear. In linear settings, LASSO remains more efficient.

Future work may extend MALD to categorical outcomes by defining local gradients over predicted class probabilities. In addition, methods such as the locally linear estimator for fused LASSO from \cite{padilla2018adaptive} could generalize our gradient-based approach beyond global linearity assumptions.

\bibliography{citations}

\newpage
\appendix

\section{Proof of Theorem 1}

\begin{theorem}[Exchangeability of Conditional Residual Knockoffs]
Let $X = (X_1,\ldots, X_p) \in \mathbb{R}^p$ be a random vector, and the observed data are $\bX \in \mathbb{R}^{n \times p}$. Denote term-wise conditional expectation predictors $ \wh{\bX}_{\cdot j} = \wh{g}_j\left(\bX_{\cdot, -j} \right)$, $j=1,2,\ldots,p$ fit on the observed data $\bX$, and assume that $\wh{g}_j$ is an unbiased estimator of $\mathbb{E}[X_j \mid X_{-j}]$. Define the residual vectors $\bGamma_{\cdot j} = \bX_{\cdot j} - \wh{\bX}_{\cdot j}$. Suppose we can generate knockoff residuals $\wt{\bGamma}_{\cdot j}$ such that
\[
    (\bGamma, \wt{\bGamma}) \stackrel{d}{=} (\bGamma, \wt{\bGamma})_{\text{swap}(S)}, \quad \forall S \subseteq \{1,\ldots,p\}.
\]
Then, defining knockoff variables by $\wt{\bX} = \wh{\bX} + \wt{\bGamma}$,
it follows that
\[
    (\bX, \wt{\bX}) \stackrel{d}{=} (\bX, \wt{\bX})_{\text{swap}(S)}, \quad \forall S \subseteq \{1,\ldots,p\}.
\]
\end{theorem}

\begin{proof}
Let $\bX \in \mathbb{R}^{n \times p}$ denote the observed data matrix, where each column $\bX_{\cdot j} \in \mathbb{R}^n$ is the vector of observed values for variable $X_j$ across $n$ samples. Let $\wh{\bX}_{\cdot j} = \wh{g}_j(\bX_{\cdot,-j})$ be the vector of predicted conditional expectations, and define the residuals:
\[
    \bGamma_{\cdot j} = \bX_{\cdot j} - \wh{\bX}_{\cdot j}, \quad \text{and} \quad \wt{\bX}_{\cdot j} = \wh{\bX}_{\cdot j} + \wt{\bGamma}_{\cdot j}.
\]
Assume the following:
\begin{itemize}
    \item[(A1)] For each $j$, the predictor $\wh{g}_j$ is fixed (i.e., deterministic) once trained on the full data $\bX$, and therefore $\wh{\bX}$ is a fixed function of $\bX$.
    \item[(A2)] The joint distribution of $(\bGamma, \wt{\bGamma})$ is invariant under coordinate-wise swapping:
    \[
        (\bGamma, \wt{\bGamma}) \stackrel{d}{=} (\bGamma, \wt{\bGamma})_{\text{swap}(S)} \quad \text{for all } S \subseteq \{1,\ldots,p\}.
    \]
\end{itemize}

We now prove the desired exchangeability property:
\[
    (\bX, \wt{\bX}) \stackrel{d}{=} (\bX, \wt{\bX})_{\text{swap}(S)} \quad \text{for all } S \subseteq \{1,\ldots,p\}.
\]

Let $\bU := \wh{\bX}$, which is a fixed matrix given $\bX$ by Assumption A1. 

Then we have:
\[
    \bX = \bU + \bGamma, \quad \wt{\bX} = \bU + \wt{\bGamma}.
\]

For any measurable subsets $\bR, \wt{\bR} \subseteq \mathbb{R}^{n \times p}$ and any subset of variables $S \subseteq \{1,\ldots,p\}$, consider the probability:
\[
\begin{aligned}
&\pr\left( [\bX, \wt{\bX}]_{\text{swap}(S)} \in [\bR, \wt{\bR}] \right) \\
=& \pr\left( [\bU + \bGamma, \bU + \wt{\bGamma}]_{\text{swap}(S)} \in [\bR, \wt{\bR}] \right) \\
=& \pr\left( [\bGamma, \wt{\bGamma}]_{\text{swap}(S)} + [\bU, \bU]_{\text{swap}(S)} \in [\bR, \wt{\bR}] \right) \\
=& \pr\left( [\bGamma, \wt{\bGamma}] + [\bU, \bU] \in [\bR, \wt{\bR}] \right) \\
=& \pr\left( [\bX, \wt{\bX}] \in [\bR, \wt{\bR}] \right),
\end{aligned}
\]
where the third equality uses that swapping a fixed quantity (\(\bU\)) has no effect, and the fourth equality uses the assumed exchangeability of $(\bGamma, \wt{\bGamma})$ from A2.

Since this equality holds for all measurable sets and all subsets \( S \), it follows that:
\[
    (\bX, \wt{\bX}) \stackrel{d}{=} (\bX, \wt{\bX})_{\text{swap}(S)}.
\]
\end{proof}

\newpage


\section{Tissue Categories for Genetic Data}\label{section:appendix-tissue-0}

All tissue types are in table \ref{tab:tissue-all-0}, and the 7 categories used for knockoffs and modeling are in table \ref{tab:tissue-reduced-0}; consolidations are informed by \cite{karlsson2021single}, \cite{kamrani2023anatomy}.

\begin{table}[ht]
    \centering
    \begin{tabular}{c|c}
        Tissue & Count\\
        \hline
        Liver & 1436\\
        Blood & 865\\
        Cortex & 612\\
        Muscle & 428\\
        Striatum & 349\\
        Cerebellum & 264\\
        Heart & 226\\
        Kidney & 179\\
        Skin & 142\\
        Spleen & 124\\
        Fibroblast & 115\\
        Brain & 91\\
        Lung & 77\\
        Tail & 70\\
        ES & 66\\
        MacrophagePeritoneal & 42\\
        Retina & 38\\
        BoneMarrow & 36\\
        WholeBrain & 34\\
        Hippocampus & 30\\
        HematopoieticStem.Progenitor.LSK & 27\\
        Heart.Atrium & 25\\
        Placenta & 24\\
        MacrophageBoneMarrow & 24\\
        Aorta & 23\\
        Hypothalamus & 20\\
        SVZ & 19\\
        Pancreas & 18\\
        Ear & 18\\
        Villi & 15\\
        Gut & 15\\
        HematopoieticStemCells & 15\\
        DifferentiatedBloodCells & 15\\
        Crypts & 15\\
        Serum & 13\\
        Adipose & 12\\
        Heart.Ventricle & 12\\
        BrainTumor & 9\\
        Astrocyte & 8\\
        Heart.Teratoma & 6\\
        iPSC & 3\\
        Pituitary & 2
    \end{tabular}
    \caption{Counts of all tissue types for mouse data.}
    \label{tab:tissue-all-0}
\end{table}

\begin{table}[ht]
    \centering
    \begin{tabular}{ l | l | l}
         Reduced Category & Original Categories & Count  \\
         \hline
         Nervous* & Cortex, Striatum, Cerebellum, Brain, Retina, & 1456\\
            & WholeBrain, Hippocampus, SVZ, BrainTumor, &\\
            & Astrocyte, Pituitary &\\
         Liver & Liver & 1422\\
         Blood* & Blood, BoneMarrow, DifferentiatedBloodCells, iPSC & 919\\
         Organ* & Kidney, Skin, Spleen, Lung, Tail, Hypothalamus,& 690\\
            & Pancreas, Ear, Villi, Gut, Crypts & \\
         Connective* & Muscle, Fibroblast, Placenta, Adipose & 576\\
         Heart* & Heart, Heart.Atrium, Aorta, Heart.Ventricle, Heart.Teratoma & 292\\
         Immune/Stem* & ES, MacrophagePeritoneal, HematopoieticStem.Progenitor.LSK, & 187\\
            & MacrophageBoneMarrow, HematopoieticStemCells, Serum & 
    \end{tabular}
    \caption{Mouse tissue categories, reduced to 7 distinct ones.}
    \label{tab:tissue-reduced-0}
\end{table}

\end{document}